\documentclass[12pt,a4paper]{article}
\usepackage{latexsym}
\usepackage[dvips]{graphics}
\usepackage{epsfig}
\usepackage{color}
\def\spa#1.#2{\left\langle#1\,#2\right\rangle}
\def\spb#1.#2{\left[#1\,#2\right]}

\begin{document}

\begin{center}
\vspace{3cm}
\begin{flushright}
AEI-2010-113
\end{flushright}\vspace{2cm}

{\bf \large INFRARED FINITE OBSERVABLES\\[0.3cm] IN  ${\cal N}=8$ SUPERGRAVITY} \vspace{3cm}

{\bf \large L. V. Bork$^{2}$, D. I. Kazakov$^{1,2}$, G. S. Vartanov$^{1,3}$, \\[0.2cm]
and A. V. Zhiboedov$^{1,4}$}\vspace{0.5cm}

{\it $^1$Bogoliubov Laboratory of Theoretical Physics, Joint
Institute for Nuclear Research, Dubna, Russia, \\
$^2$Institute for Theoretical and Experimental Physics, Moscow, Russia, \\
$^3$Max-Planck-Institut f\"ur Gravitationsphysik,
Albert-Einstein-Institut 14476 Golm, Germany,\\ $^4$Department of
Physics, Princeton University, Princeton, NJ 08544,
USA.}\vspace{1cm}

\begin{abstract}Using the algorithm of constructing the IR finite observables
suggested in \cite{Bork:2009ps} and discussed in details in
\cite{Bork:2009nc}, we consider construction of such observables in
${\cal N}=8$ SUGRA in NLO of PT. In general, contrary to the amplitudes defined
in the presence of some IR regulator, such observables do not reveal
any simple structure.
\end{abstract}
\end{center}

\newpage

\section{Introduction}
In the last decade remarkable progress in understanding the structure of the amplitudes in supersymmetric gauge theories has been achieved
\cite{Dixon:2008tu}. Due to development of the so-called unitarity cut technique \cite{Bern:1994cg,Bern:1994zx}, the three- and four-loop
results for the four-point amplitudes became available for $\mathcal{N}=4$ super Yang-Mils theory (SYM) \cite{4-loop}. The application
of the same technique in $\mathcal{N}=8$ supergravity (SUGRA), first used in \cite{Dunbar:1995ed,Bern:1998sv,Bern:1998ug}, made it possible
the computation of the three-  \cite{Bern:2007hh} and four-loop \cite{Bern:2009kd} four-point amplitudes in $\mathcal{N}=8$ SUGRA.
We want to stress that obtaining such results, using the standard diagram technique (component or superspace) seems
practically impossible due to extreme complexity of such computations.

This results initiated once again the discussion of possible ultraviolet (UV) finiteness of the $\mathcal{N}=8$ SUGRA \cite{SUGRA1} and 
motivated the search of possible constraints on the UV divergences in $\mathcal{N}=8$ SUGRA from various points of view
\cite{Berkovits:2006vc,Green:2006yu,Bossard:2009mn,Green:2010sp} and  of the presence of the symmetry group $E_{7(7)}$
in the theory \cite{Brink:2008qc,Kallosh:2008ic}.

The answers for the amplitudes, which can be obtained from the unitarity cut technique, are usually represented as a sum of the master scalar
integrals, defined in $D$ dimensions with fixed coefficients. This provides the possibility to analyze the dependence (UV behavior) of the amplitude
on the dimensionality of space-time $D$. Such an analysis up to four loops for the four-point amplitudes in $\mathcal{N}=8$ SUGRA revealed the same
UV behavior of the amplitudes as in $\mathcal{N}=4$ SYM. This result, as claimed by the authors of \cite{SUGRA1,SUGRA2}, does not follow from
the properties of $\mathcal{N}=8$ SUSY and cannot be obtained, at least in some obvious way, from the previous analysis based on different
versions of superspace technique (see, however,  \cite{SUGRA3}). This makes it possible, once again, to suggest that 
$\mathcal{N}=8$ SUGRA amplitudes (the S-matrix) are \cite{SUGRA1,SUGRA2} UV finite as in the case of $\mathcal{N}=4$ SYM.

The expressions obtained by means of the unitarity cut technique in $\mathcal{N}=4$ SYM and $\mathcal{N}=8$ SUGRA usually have a very "simple" form. This, and possible
UV finiteness of the $\mathcal{N}=8$ SUGRA, allows one to make a conjecture, that the amplitudes (the S-matrix) in $\mathcal{N}=8$ SUGRA are the "simplest"  among all $D=4$-dimensional QFTs with maximum spin $\leq2$\cite{SimpleQFT}.

However, despite  the compact and simple form and the UV finiteness, the amplitudes in $\mathcal{N}=8$ SUGRA (as in any $D=4$ dimensional QFT
with massless fields) are still, strictly speaking, ill defined in $D=4$ (without  some infrared (IR) regulator) due to the presence of the IR divergences.

In the case of Yang-Mills (YM) theories (and QCD) in such cases the objects of physical interest are not the amplitudes themselves but the
so-called IR finite observables which are build from the amplitudes, but possess no dependence on the IR regulator. In
\cite{Bork:2009ps,Bork:2009nc}  the authors suggested the example of such observables for $\mathcal{N}=4$ SYM. In contrast to relatively
simple answers for several first orders of PT for the four- and five-point amplitudes \cite{BDS}, the IR finite observables considered in \cite{Bork:2009ps,Bork:2009nc}, in the first nontrivial order of PT (NLO) has no simple structure and resemble the
complicated (prior to properly applied simplifications) expressions for the six-point amplitudes in the second order of PT \cite{DelDuca:2009au}.

It seems natural to consider  the similar IR finite observables  in $\mathcal{N}=8$ SUGRA. It is this
kind of observables that have a clear physical meaning. In the current article we discuss the construction of such observables in
$\mathcal{N}=8$ SUGRA using the same methods as in \cite{Bork:2009ps,Bork:2009nc}.

\section{Inclusive cross-sections as IR finite observables}

One of the possible IR finite observables are properly constructed perturbative inclusive cross-sections. Such observables naturally
appear in the parton model of perturbative QCD \cite{EKSglu,EKS,KunsztSoper,Kunszt3Jet,Katani} and their
construction is based on the Kinoshita-Lee-Nauenberg (KLN) theorem \cite{KLN}. In such inclusive cross-sections, if the dimensional
regularization is used, the IR divergences appears as $1/\epsilon$ poles. Cancellation of this  infrared divergences coming from the loops
includes two main ingredients:  emission of additional soft real quanta and redefinition of the asymptotic states resulting in the
splitting terms governed by the Altarelli-Parisi kernels. The latter ones take care of the collinear divergences and 
are absorbed into the probability distributions of initial and final particles over the fraction of momenta $q_i(z)$. The appearance of such 
distributions can be heuristically understood in the following fashion:  in massless QFT a particle with momentum $p$ is indistinguishable from 
the jet of particles with the same overall momentum and quantum numbers flying parallell. In the  first none trivial order of PT in dimensional 
regularization the momentum distribution $q_i(z)$ can be represented as
\begin{eqnarray}\label{q}
q_i(z,\frac{Q_{f}^2}{\mu^2})&=&\delta(1-z)+\frac{g^2}{2\pi}\frac{1}{\epsilon}
\left(\frac{\mu^2}{Q_{f}^2}\right)^\epsilon \sum_j P_{ij}(z),
\end{eqnarray}
where $P_{ij}(z)$ are the so-called splitting functions. They define the probability  that  the particle $i$ emits the  collinear particle $j$ with a fraction of momentum $z$. Here $g$ is the coupling constant, $\mu$ is the mass parameter of dimensional regularization.
$Q_f$ is usually called the factorization scale and can be interpreted as the width of a jet of collinear particles. The contributions from
$P_{ij}(z)$ to the cross-section sometimes are called the collinear counterterms.

Schematically, the class of IR finite observables discussed above, which are the inclusive cross-sections,  can be written as
\begin{eqnarray}\label{IRfinGen}
d\sigma^{incl}_{obs}&=&\sum\limits_{n=2}^{\infty}\int\limits_0^1\!dz_1q_1(z_1,
\frac{Q_f^2}{\mu^2})\! \int\limits_0^1\!dz_2q_2(z_2,
\frac{Q_f^2}{\mu^2})\prod\limits_{i=1}^{n}\int\limits_0^1\!dx_iq_i(x_i,
\frac{Q_f^2}{\mu^2})\times \\
   &&\hspace{-1cm}\times  d\sigma^{2\to n}(z_1p_1,z_2p_2,...)S_n(\{z\},\{x\})=
   g^4\sum\limits_{L=0}^\infty \ \left(\frac{g^2}{16\pi^2}\right)^L d\sigma_L^{Finite}
   (s,t,u,Q_f^2), \nonumber
\end{eqnarray}
where $\mathcal{S}_n(\{z\},\{x\})$ are the so-called measurement functions, which define what inclusive cross-section (total, differential
distribution, etc.) is considered. Note that not for any choice of $\mathcal{S}_n(\{z\},\{x\})$ one gets the IR finite result (see
\cite{Bork:2009ps,Bork:2009nc} for discussion).

\section{Computation of inclusive cross-section in ${\cal N}=8$ SUGRA}
Our aim is to evaluate the inclusive differential polarized cross-section in NLO in the week coupling limit in planar ${\cal N}=8$ SUGRA  
in analytical form and to trace the cancelation of the IR divergences. 

Consider the inclusive cross-section for the scattering of polarized 
gravitons in $\mathcal{N}=8$ SUGRA in the first nontrivial order of PT and follow the algorithm discussed in \cite{Bork:2009ps,Bork:2009nc}.
We  assume that all the gravitons have a fixed helicity $(+)$ or $(-)$ and consider all of them to be outgoing ones.  The incoming graviton  has the 
opposite momentum and  helicity.  Then all the tree amplitudes with helicity configurations $(+...+)$ or $(-+...+)$ are identically zero. The first nonzero tree level graviton amplitudes, similar to the YM case, are  those with helicity configurations $(--+...+)$. They  are called the $\mbox{MHV}$ (or
$\mbox{anti-MHV}$ in the case $(++-...-)$) amplitudes. We restrict ourselves in our consideration to the class of the $\mbox{MHV}$ amplitudes 
(the $\mbox{MHV}$ channel) only.  Consideration of the $\mbox{anti-MHV}$ channel can be done in the similar fashion. Note, however, that  such a restriction  is possible only in the leading order.  In higher orders of PT  in order to achieve the IR finiteness, in general,  one has to consider all type of the amplitudes since they are mixed.

Consider the process of   $2\times 2$ graviton scattering where all incoming and all outgoing gravitons have positive helicity.  In our notations this corresponds to the amplitude with  helicity configuration $(--++)$. This amplitude is the $\mbox{MHV}$ and $\mbox{anti-MHV}$ at the same time. 
So one has to take it with the weight $1/2$ for considering the $\mbox{MHV}$ channel only.

The tree amplitude under consideration  written in helicity spinor formalism \cite{Dixon}  has
the form
\begin{equation}{\cal M}_4^{tree}(1^-, 2^-, 3^+, 4^+)  =(16 \pi G_{N}) i
\, \spa1.2^8 { \spb1.2 \over  \spa3.4 \, N(4) } ,
\label{GravTreeFour}
\end{equation}
where
\begin{equation}
 N(n) \equiv \prod_{i=1}^{n-1} \prod_{j=i+1}^n \spa{i}.{j} 
\label{NnDef}.
\end{equation}
Here the notation, now standard, for the spinor inner product has been used:
\begin{eqnarray}
\epsilon^{ab}\lambda_{a}^{(i)}\lambda_{b}^{(j)}=\langle\lambda^{(i)}\lambda^{(j)}\rangle
\doteq \langle ij\rangle,
~~~\epsilon^{\dot{a}\dot{b}}\bar{\lambda}_{\dot{a}}^{(i)}\bar{\lambda}_{\dot{b}}^{(j)}
=[\bar{\lambda}^{(i)}\bar{\lambda}^{(j)}] \doteq [ ij].
\end{eqnarray}
Under the complex conjugation one has
\begin{equation}
(\langle ij\rangle)^{*}=[ ij],
\end{equation}
and also
\begin{equation}
\langle ij\rangle[ij]=s_{ij},
\end{equation}
where $s_{ij}=(p_{i}+p_{j})^2$, and $p_{i},~p_{j}$ are some on-shell momenta which correspond to the $i$-th and $j$-th particles.

Then the differential cross-section $d\sigma(g^-g^-\to g^+g^+)/d\Omega$ can be written as
\begin{eqnarray}\label{tree}
  \left(\frac{d\sigma}{d\Omega_{13}}\right)_0^{(--++)} = \frac{1}{E^2} \int d\phi_{2} | {\cal
 M}_{4}^{tree|}|^2\mathcal{S}_2,
\end{eqnarray}
where $d\phi_{2}$  is the two particle phase space,
$\mathcal{S}_n$  is the measurement function which, in our case, has
the form
\begin{eqnarray}
\mathcal{S}_2=\delta_{+,h_{3}} \delta^{D-2} (\Omega - \Omega_{13}),
\end{eqnarray}
$d\Omega_{13}=d\phi_{13}d cos(\theta_{13})$, $\theta_{13}$ is the scattering angle of a particle with momentum $p_3$ with respect to the
particle with momentum $p_1$ in the center of mass (c.m.) frame, $\delta_{+,h_{3}}$ corresponds to the fact that particle with
momentum $p_3$ has the positive helicity.

From now on the dimensional regularization (reduction) will be used. In the c.m. frame the Born contribution to the cross-section can then be written as
\begin{eqnarray} &&
\hspace{-1.5cm}\left(\frac{d\sigma}{d\Omega}\right)_0^{(--++)}\hspace{-0.3cm}
=\frac{1}{E^2}
 \frac{\alpha_{Gr}^2 s^6}{t^2 u^2}\left(\frac{\mu^2}{s}\right)^\epsilon
=\frac{(\alpha_{Gr}
E^2)^2}{E^2}\left(\frac{\mu^2}{s}\right)^\epsilon\frac{16}{(1-c^2)^2},
 \end{eqnarray}
where $s,t,u$ are the standard Mandelstam variables, $E$ is the total energy of initial particles in the c.m. frame, $c=\cos \theta_{13}$
is the cosine of the scattering angle of the third particle, $\mu$ and $\epsilon$ are the parameters of
dimensional regularization, and $\alpha_{Gr}=G_{N}/4 \pi$. In the c.m. frame for the Mandelstam variables one has
$$s=E^2,\ \  t=-E^2/2(1-c), \ \  u=-E^2/2(1+c).$$

\subsection{Virtual contribution}
Consider now the one-loop correction to the Born contribution. The corresponding amplitude has the form:
\begin{eqnarray} && {\cal M}_4^{\mathcal{N}=8}(1^-, 2^-, 3^+, 4^+)  = (16 \pi
G_{N})^2 {1\over 4} \spa1.2^8 \nonumber \\ && \makebox[2em]{} \times
\, \Bigl[ h(1,\{2\},3) h(3,\{4\},1) Tr^2[1234] \, I_4^{1234} +
\hbox{perms}\Bigr], \label{N8Explicit} 
\end{eqnarray}
 where $Tr[i_1 i_2 i_3
i_4] \equiv Tr[ \hat k_{i_1} \hat k_{i_2} \hat k_{i_3} \hat k_{i_4}]$ and 
$$
h(a,\{1\},b) = {1\over \spa{a}.{1}^2 \spa{1}.{b}^2} \,.
\label{Half1}$$
The summation goes over all possible permutations  and the
scalar integral $I_4^{1234}$ is equal to
$$ I_4^{1234}(s,t) \ = \
-\frac{2}{st}\frac{\Gamma(1+\epsilon)\Gamma(1-\epsilon)^2}{\Gamma(1-2\epsilon)}
\left(\frac{1}{\epsilon^2}(\left(\frac{\mu}{s}\right)^{\epsilon}+\left(\frac{\mu}{-t}\right)^{\epsilon})
+\frac{1}{2}\log^2\left(\frac{s}{-t}\right)+\frac{\pi^2}{2}\right)+\mathcal{O}(\epsilon).
$$
So the contribution to the cross-section  in c.m. frame  has the form
\begin{eqnarray}\label{1loop4gluon}
\left(\frac{d\sigma}{d\Omega}\right)_{virt}^{(--++)}&=&\frac{(\alpha_{Gr}
E^2)^3}{\pi E^2}\left(\frac{\mu^2}{s}\right)^{2\epsilon}
\frac{64}{(1-c^2)^2} \left[\frac{1}{\epsilon} \left((1+c)\log
(\frac{1+c}{2}) \right. \right. \\ \nonumber & &\left. \left.
+(1-c)\log (\frac{1-c}{2})\right) + 2 \log (\frac{1+c}{2}) \log
(\frac{1-c}{2})\right].
\end{eqnarray}
It should be stressed that   in this order of PT the UV divergences in  $\mathcal{N}=8$ are absent and all the divergences have the IR nature. Note
also the absence of the $1/\epsilon^2$ pole which in this case cancels due to permutations in (\ref{N8Explicit}) contrary to the gauge theories  in $D=4$
where it is usually present.   As one can see from (\ref{1loop4gluon}), despite the  extremely complicated intermediate expressions  appearing  in diagrammatical computations, the final result for the four-point one-loop $\mbox{MHV}$ graviton amplitude in $\mathcal{N}=8$ has a relatively "simple" form.

\subsection{Real emission}

Following the algorithm for construction of the IR finite observables  we consider the contribution to the cross-section 
from amplitudes with additional particles in the final state. For the fixed helicity of initial particles one has two types of the graviton
amplitudes:
\begin{enumerate}
  \item Three gravitons in the final state with positive helicity: $g^-g^- \rightarrow g^+g^+g^+.$
  This is the $\mbox{MHV}$ amplitude;
  \item Two gravitons in the final state with positive and one with
  negative helicity: $g^-g^- \rightarrow  g^+g^+g^-.$
  This is the $\mbox{anti-MHV}$ amplitude.
\end{enumerate}

In the $\mbox{MHV}$ channel only the first amplitude contributes. In this case there are three identical particles in the final state so we have 
to choose which particle we are detecting. This can be achieved, for example,  by arranging the particles in according  to their energy and
selecting "the fastest one". It would correspond to the measurement function of the form
\begin{eqnarray}\label{mf3p1a}
{\cal S}^{(--+++),1}_{3} = 
\delta_{+,h_{3}} \Theta(p^0_{3} > p^0_{4}) \Theta(p^0_{3} >
p^0_{5})\delta^{D-2} (\Omega - \Omega_{13}).
\end{eqnarray}
In practice it is more convenient to work with the measurement function written as \cite{Bork:2009nc}
\begin{equation}\label{mf} \mathcal{S}_3(p_3,p_4,p_5) =
 \Theta(p_3^0 - \frac{1 - \delta}{2} E)
\delta^{D-2}(\Omega-\Omega_{3}),
\end{equation}
where $\delta$ is an arbitrary parameter which can be fixed, for example,  from the requirement that the detected particle is the
fastest one (this corresponds to $\delta=1/3$). We will leave the value of  $\delta$ arbitrary to be convicted  that the cancelation of the  IR
divergences occurs for arbitrary $\delta$.

The amplitude for the process  $g^-g^- \rightarrow g^+g^+g^+$ can be written as:
\begin{equation}
 {\cal M}_5^{tree}(1^-, 2^-, 3^+, 4^+, 5^+) =(16 \pi
G_{N})^{\frac{3}{2}} i \, \spa1.2^8 { \epsilon(1,2,3,4) \over N(5) },  \label{GravTreeFourFive}
\end{equation}
where
$$
\epsilon(i,j,m,n) \equiv 4i\epsilon_{\mu\nu\rho\sigma} k_i^\mu k_j^\nu
k_m^\rho k_n^\sigma \ =\
\spb{i}.{j}\spa{j}.{m}\spb{m}.{n}\spa{n}.{i}
   - \spa{i}.{j}\spb{j}.{m}\spa{m}.{n}\spb{n}.{i} \, .
\label{LeviCivitaDef}
$$

Then, the cross-section is
\begin{eqnarray}\label{crossf}
 \left(\frac{d\sigma}{d\Omega_{13}}\right)^{(--+++)}_{Real} =  \frac{1}{E^2} \int d\phi_{3} | {\cal
 M}_5|^2 \mathcal{S}_3,
\end{eqnarray}
where $d\phi_{3}$ is three particle phase space.  After the integration one has
\begin{eqnarray}
&&\hspace{-0.5cm}\left(\frac{d\sigma}{d\Omega_{13}}\right)^{(--+++)}_{Real}
=\frac{(\alpha_{Gr} E^2)^3}{\pi
E^2}\left(\frac{\mu^2}{s}\right)^{2\epsilon} \frac{64}{(1-c^2)^2}
\left[\frac{1}{\epsilon} \left((1+c)\log (\frac{1+c}{2}) \right.
\right. \\ \nonumber & &\left. \left.
+(1-c)\log (\frac{1-c}{2})\right) 
+\mbox{Finite part}(\delta,c) \right].
\end{eqnarray}
The finite part is a complicated polynomial function of  $\log$, $\log^2$ and $Li_2$ with the argument of the form $(1 \pm c)/2$ and in
general have the same structure as in \cite{Bork:2009nc}\footnote{One can get it upon request from the authors}.

\subsection{Collinear counterterms}

Consider now the  contributions from collinear counterterms. In the $\mbox{MHV}$ channel the splitting function $P_{ij}(z)$ has the form
(we use here slightly different notation for the splitting functions indicating explicitly all three particles like
$P^{init}_{fin_1,fin_2}(z)$ to avoid confusion.)
\begin{equation}
P^{g^-}_{g^+g^+} = [\frac{1}{z (1-z)_{+}}]^2 =\frac{1}{z^2} +
\frac{2}{z (1-z)_{+}} + \frac{1}{[(1-z)^2]_{+}}.
\end{equation}
Here $1/[(1-z)^2]_{+}$ should be understood as
$$
\int dz \frac{f(z)}{[(1-z)^2]_{+}} = \int dz
\frac{f(z)-f(1)-f'(1)(z-1)}{(1-z)^2}.
$$

The splitting function $P^{g^-}_{g^+g^+}$ can be obtained as the collinear limit of the corresponding amplitude \cite{Dixon}. Note
also that in the case of $\mathcal{N}=8$ SUGRA the probability distribution $q(z)$ does not receive radiative corrections in $\alpha_{Gr}$ except  for the
first order\cite{Bern:1998sv}.

The contribution to the cross-section from the IR counterterms can be schematically written as
\begin{eqnarray}\label{AP}
d \sigma_{2 \rightarrow 2}^{spl,init} =
\frac{\alpha_{Gr}}{2\pi}\frac{1}{\epsilon}
\left(\frac{\mu^2}{Q^2_f}\right)^\epsilon \int_{0}^{1} d z
P^{g^-}_{g^+g^+}(z)\sum\limits_{i,j=1,2;\ i\neq j}d \sigma_{2
\rightarrow 2} (z p_{i},p_{j},p_{3},p_{4}){\cal S}_2^{spl,init}(z),
\end{eqnarray}
\begin{eqnarray}\label{APfin}
d \sigma_{2 \rightarrow 2}^{spl,fin} =
\frac{\alpha_{Gr}}{2\pi}\frac{1}{\epsilon}
\left(\frac{\mu^2}{Q^2_f}\right)^\epsilon d \sigma_{2 \rightarrow 2}
(p_{1},p_{2},p_{3},p_{4}) \int_{0}^{1} d z P^{g^-}_{g^+g^+}(z) {\cal
S}_2^{spl,fin}(z).
\end{eqnarray}

The measurement function in this case has the same form  as in the case of the real emission, but now depends on the fraction of
momentum $z$ that restricts the integration over $z$
\begin{equation}\label{sz}
{\cal S}_2^{spl,1}(z) = \delta_{+,h_3}
  \delta^{D-2} (\Omega-\Omega_{13})
  \theta(z-z_{min}),
\end{equation}
where $z_{min}$  is equal to
\begin{equation}\label{Up1}
z_{min}^{in} \ = \ \frac{(1 - \delta) (1 - c)}{1 + \delta - c (1 -
\delta)}, \ \ \ z_{min}^{fin} \ = \ (1 - \delta)
\end{equation}
for the splitting function of the  initial and final states, respectively. This conditions can be obtained from the requirement  $p^0_3>(1-\delta)E/2$ in the appropriate kinematics.

Integration over the phase space and over the fraction of momentum $z$ gives:
\begin{eqnarray}
&&\left(\frac{d\sigma}{d\Omega_{13}}\right)^{(--+++)}_{InSplit} =
\frac{(\alpha_{Gr} E^2)^3}{\pi
E^2}\left(\frac{\mu^2}{s}\right)^{2\epsilon} \frac{128}{(1-c^2)^2}
\left[ \frac 1\epsilon \left(\frac{1-2 \delta}{(\delta-1) \delta} -
2 \log (1-\delta) \right. \right.\\ \nonumber && \left. \left.+ 2
\log\delta - (1-c) \log\frac{1-c}{2} - (1+c) \log\frac{1+c}{2}
\right)+\mbox{Finite part}(\delta,c)  \right],
\end{eqnarray}
\begin{equation}
\left(\frac{d\sigma}{d\Omega_{13}}\right)^{(--+++)}_{FnSplit} =
\frac{(\alpha_{Gr} E^2)^3}{\pi
E^2}\left(\frac{\mu^2}{s}\right)^{2\epsilon} \frac{128}{(1-c^2)^2}
\frac 1\epsilon \left[ \frac{2 \delta-1}{(\delta-1) \delta} + 2
\log(1-\delta) - 2 \log (\delta) \right].
\end{equation}

\section{IR finite observables in ${\cal N}=8$ SUGRA}

It is easy to see that the \mbox{$\mbox{MHV}$} part of inclusive cross-section defined  in (\ref{IRfinGen}) as a  sum of several contributions
\begin{equation} \label{fin1} A^{MHV}=\frac12 \left(\!\frac{d\sigma}{d\Omega_{13}}\!\right)^{\index
 (--++)}_{0}\hspace{-0.3cm}+\frac
12\left(\!\frac{d\sigma}{d\Omega_{13}}\!\right)^{(--++)}_{Virt}\hspace{-0.3cm}
+\left(\!\frac{d\sigma}{d\Omega_{13}}\!\right)^{(--+++)}_{Real}\hspace{-0.3cm}+
\left(\!\frac{d\sigma}{d\Omega_{13}}\!\right)^{(--+++)}_{InSplit}\hspace{-0.3cm}+
\left(\!\frac{d\sigma}{d\Omega_{13}}\!\right)^{(--+++)}_{FnSplit}\!\!\!\!,
\end{equation}
is IR finite, all the divergences cancel as expected. It should be stressed that this cancellation occurs for arbitrary $\delta$.  The analogous 
cancellation should take place in  the $\mbox{anti-MHV}$ channel. Note that  $A^{MHV}$ has no simple structure in contrast
to the virtual correction to the cross-section $(d\sigma/d\Omega_{13})^{(--++)}_{0}$.

\section{Conclusion}

We have applied the algorithm for construction of  IR finite observables  in $\mathcal{N}=8$ SUGRA and explicitly demonstrated the
cancellation of the IR divergences in NLO of PT. One can see that such observables, in general, have no any
simple structure in contrast to the amplitudes defined in the presence of some IR regulator. 
In \cite{Bork:2009ps,Bork:2009nc} a similar analysis has been performed for  the $\mathcal{N}=4$ SYM theory with the same conclusions.
This makes us to conclude that this type of observables is not optimal in the sense of "simplicity" of the answer. 
It is also important to note the dependence of the answers on the factorization scale  $Q_f$ which breaks the conformal invariance. This dependence is a general feature of observables constructed on the base of eq.(\ref{IRfinGen}).

At the same time there are no "no-go" theorems which prohibit the existence of observables reflecting  the rich symmetries of maximally supersymmetric YM or gravity theories. The search of such observables seems to be an interesting challenge, if one wants to construct physically meaningful expressions which reveal the possible integrability structure of the model.

\section*{Acknowledgements}

Financial support from RFBR grant \# 08-02-00856 and grant of the Ministry of Education and Science of the Russian Federation \# 1027.2008.2 is kindly acknowledged.

\end{document}